\documentclass{iau} 
\usepackage{graphicx}
\usepackage{aas_macros}
\title[Clustered star formation and (proto)planetary systems] 
{The implications of clustered star formation \\ for (proto)planetary systems and habitability}

\author[J.~M.~Diederik Kruijssen \& Steven~N.~Longmore]
{J.~M.~Diederik Kruijssen$^1$
 \and Steven~N.~Longmore$^2$}

\affiliation{$^1$Astronomisches Rechen-Institut, Zentrum f\"{u}r Astronomie der Universit\"{a}t Heidelberg,\\ M\"{o}nchhofstra\ss e 12-14, 69120 Heidelberg, Germany \\ email: {\tt kruijssen@uni-heidelberg.de} \\[\affilskip]
$^2$Astrophysics Research Institute, Liverpool John Moores University,\\ IC2, Liverpool Science Park, 146 Brownlow Hill,
Liverpool L3 5RF, United Kingdom \\email: {\tt s.n.longmore@ljmu.ac.uk}}

\pubyear{2019}
\volume{345}  
\jname{Origins: from the Protosun to the First Steps of Life}
\editors{Bruce G. Elmegreen, L. Viktor T\'oth, Manuel G\"udel, eds.}

\begin{document}

\maketitle

\begin{abstract}
Star formation is spatially clustered across a range of environments, from dense stellar clusters to unbound associations. As a result, radiative or dynamical interactions with neighbouring stars disrupt (proto)planetary systems and limit their radii, leaving a lasting impact on their potential habitability. In the solar neighbourhood, we find that the vast majority of stars form in unbound associations, such that the interaction of (proto)planetary systems with neighbouring stars is limited to the densest sub-regions. However, the fraction of star formation occurring in compact clusters was considerably higher in the past, peaking at $\sim50\%$ in the young Milky Way at redshift $z\sim2$. These results demonstrate that the large-scale star formation environment affects the demographics of planetary systems and the occupation of the habitable zone. We show that planet formation is governed by multi-scale physics, in which Mpc-scale events such as galaxy mergers affect the AU-scale properties of (proto)planetary systems.
\end{abstract}

\firstsection

\section{Introduction}
The formation of stars and (proto)planetary systems proceeds in a hierarchically structured way (e.g.~Elmegreen 2008; Kruijssen 2012; Longmore et al. 2014). In the bound substructures, dynamical interactions and external photoionisation are common processes that may disrupt (proto)planetary systems and affect the occupation of the habitable zone within these systems (e.g.~Scally \& Clarke 2001; de Juan Ovelar et al.~2012). These external processes become more efficient towards higher gas and stellar densities.

Observations probing the evolution of star formation in galaxies over cosmic time show that the cosmic star formation rate peaked at redshift $z=1$--$3$ (Madau \& Dickinson 2014). On the size scale of giant molecular clumps in which the stars are born ($0.1$--$1$~kpc, Genzel et al.~2011), the interstellar medium in these galaxies is extreme when compared to the conditions in the solar neighbourhood. The gas density, gas pressure, cosmic ray ionisation rate and FUV interstellar radiation field are several orders of magnitude higher than for clouds forming stars in the vicinity of the Sun (Swinbank et al.~2011; Tacconi et al.~2013). This means that the conditions under which most stars and planetary systems formed were likely very different than the conditions under which we observe the formation of planetary systems today, with plausibly important implications for the demographics of the planet population.

\section{Environmental effects on protoplanetary discs}

The pre-ALMA suite of protoplanetary disc (PPD) observations available in the literature already demonstrated that PPD properties change as a function of the local stellar environment. Specifically, the disc radii are statistically smaller in high-density environments, dropping below 200 AU for stellar number densities $>5\times10^3$~pc$^{-2}$ (de Juan Ovelar et al.~2012). Since then, ALMA has revealed this effect in much higher detail (Eisner et al.~2018). In de Juan Ovelar et al.~(2012), we used a simple analytical model to show that the observed truncation may be caused by dynamical perturbations from passing stars, but more recent work shows that external photoevaporation is likely responsible (see below). Importantly, dynamical truncations continue for planetary systems, beyond PPD dispersal. As a result, the time for which the habitable zone may be occupied by planets (the `HZ lifetime') is a function of stellar mass and ambient stellar density. Even in a low-density cluster with a stellar density $>\{10^2,10^3\}$~pc$^{-2}$ (or $M>\{10^3,10^4\}$~M$_\odot$), Earth would have been ejected from the habitable zone by the present day, with HZ lifetimes of $<\{4,0.3\}$~Gyr, respectively (de Juan Ovelar et al.~2012).

Rosotti et al.~(2014) present hydrodynamical simulations of 50 PPDs in a bound stellar cluster. The PPDs evolve self-consistently due to viscous spreading and dynamical perturbations. They undergo tidally-driven morphological transformations, which in extreme cases can completely destroy disc in a few~$10^5$~yr. The PPD radii are set by closest dynamical encounter, where disc shrinkage occurs for closest encounter distances less than $\sim6$ disc radii. Consistently with the analysis by de Juan Ovelar et al.~(2012), the simulations show that PPDs in high-density environments are statistically smaller, dropping below 200~AU for projected stellar number densities $>3\times10^3$~pc$^{-2}$.

All observed candidates of environmentally-driven shrinkage in the de Juan Ovelar et al.~(2012) sample are located in the Orion Nebula Cluster. These discs are well known to be affected by external photoevaporation (Henney \& O'Dell 1999; Scally \& Clarke 2001). The question thus rises whether disc disruption is dominated by dynamical encounters or photoevaporation. Winter et al.~(2018) model disc destruction by both processes and show that external photoevaporation dominates in clusters hosting massive stars (affecting PPDs for UV flux densities of $>3\times10^3$~G$_0$ or densities $>20$~pc$^{-3}$). Dynamical encounters dominate only in low-mass regions without massive stars and require extreme densities of $>10^4$~pc$^{-3}$. This shows that external UV irradiation generally dominates over PPD dispersal by dynamical encounters, and that (proto)planetary systems in (bound) clusters are systematically subjected to externally-driven disruption.

\section{The fraction of planetary systems born in dense stellar clusters}

A wide range of recent work shows that the efficiency of cluster formation depends on the local gas density (Elmegreen 2008; Kruijssen 2012). In the interstellar medium, the gas density probability distribution function (PDF) is lognormal, of which the width increases with the gas pressure (Vazquez-Semadeni 1994; Padoan et al.~1997; Krumholz \& McKee 2005). Towards the high-density tail of the density PDF, the gas has short free-fall times. At a fixed star formation efficiency per free-fall time (e.g.~Leroy et al.~2017; Utomo et al.~2018), this implies high integrated star formation efficiencies, low gas fractions, and, as a result, high bound stellar fractions (Kruijssen 2012). In other words, unbound associations form at the low-density end of the gas density PDF, whereas compact, bound stellar clusters form at the high-density end. Note that the external photoevaporation of PPDs does not strictly require the young stars to reside in a gravitationally {\it bound} cluster, but the long-term disruption of planetary systems by dynamical perturbations does.\looseness=-1

When integrating the gas density PDF, we obtain the fraction of star formation occurring in bound clusters (the cluster formation efficiency or CFE; Bastian 2008; Kruijssen 2012) as a function of the gas pressure. The resulting CFEs range from a few \% at low gas pressures (or surface densities, $\sim10$~M$_\odot$~pc$^{-2}$), such that few planetary systems are affected by environmental effects, to up to $\sim50\%$ at high gas pressures (or surface densities, $>200$~M$_\odot$~pc$^{-2}$), where many planetary systems are affected by environmental effects. The prediction that the CFE is a function of the gas pressure has been quantitatively confirmed in observational studies of young stellar cluster populations (Adamo et al.~2015; Johnson et al.~2016; Ward \& Kruijssen 2018). For instance, the solar neighbourhood is observed (Lada \& Lada 2003) and predicted (Kruijssen 2012) to form 7\% of all stars in bound stellar clusters, whereas the CFE is predicted and observed to be about 40\% at the high gas pressures near the Galactic Centre (Ginsburg \& Kruijssen 2018), which are similar to those at high redshift (Kruijssen \& Longmore 2013).

\section{Cosmological context}

Given the environmental dependence of the CFE, a general statement regarding its impact on the externally-driven dispersal of (proto)planetary systems requires a representative census of the CFE as a function of galactic environment and cosmic time. This is now possible thanks to the E-MOSAICS project (for MOdelling Star cluster population Assembly In Cosmological Simulations within EAGLE; Pfeffer et al.~2018; Kruijssen et al.~2018), which is a set of hydrodynamical cosmological `zoom-in' simulations of 25 Milky Way-mass galaxies that include a physically-motivated, sub-grid model for the formation and dynamical evolution of the entire star cluster population. These simulations represent the first time that the formation and evolution of the cluster population can be followed self-consistently across cosmic history.

\begin{figure}[b]
\vspace*{-0.5 cm}
\begin{center}
\includegraphics[width=5in]{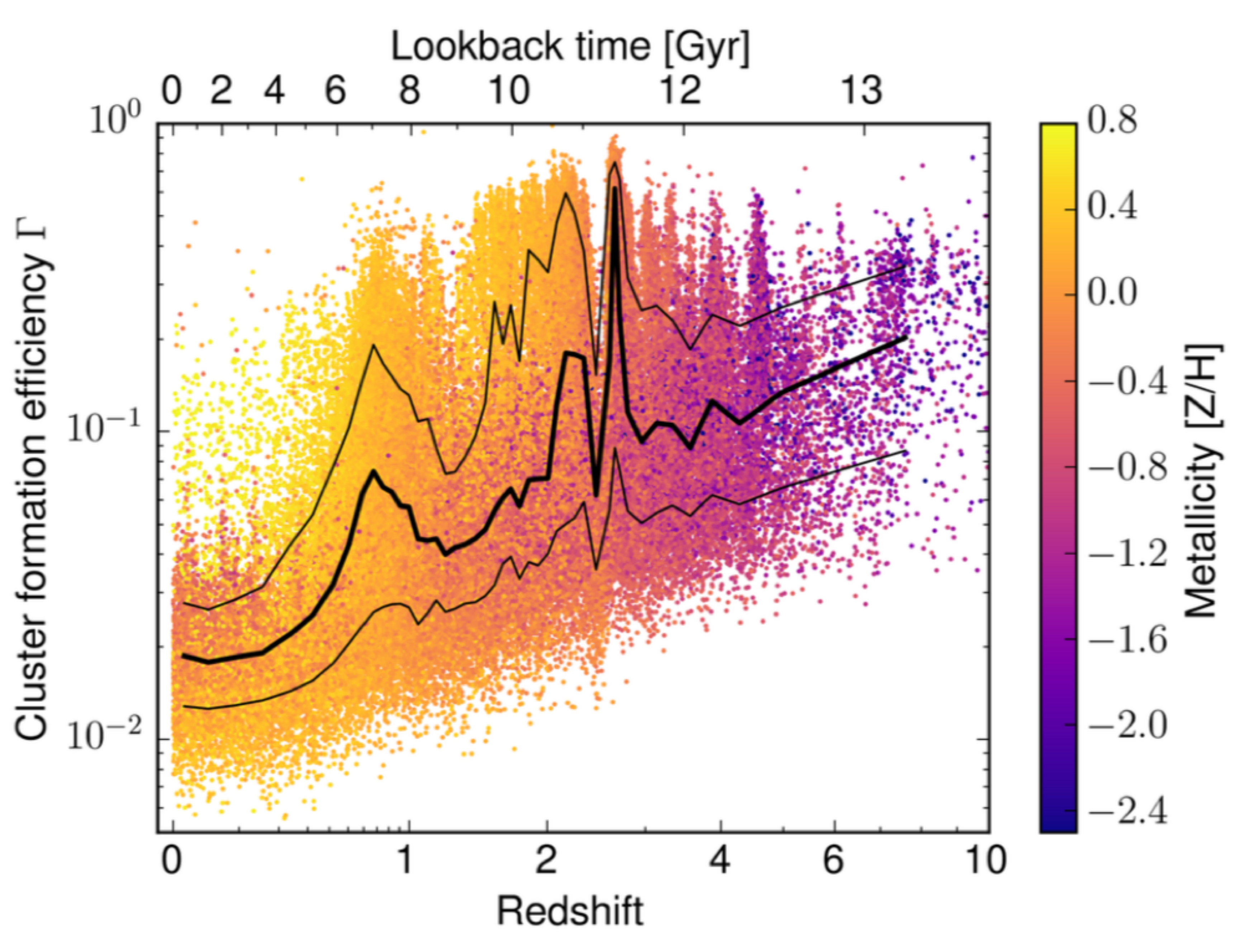} 
\vspace*{-0.3 cm}
 \caption{Cluster formation efficiency (fraction of star formation occurring in bound clusters) as a function of redshift for one of the cosmological zoom-in simulations of Milky Way-mass galaxies from the E-MOSAICS project (Pfeffer et al.~2018; Kruijssen et al.~2018), which include a model for the formation and evolution of the stellar cluster population. The data are colour-coded by metallicity, with the median and standard deviation at each redshift indicated by the thick and thin solid lines, respectively. The peaks at $z=\{0.7,2,2.5\}$ correspond to galaxy mergers, which drive an increase of the gas pressure and a corresponding increase of the CFE. As a result, the external disruption of (proto)planetary systems is enhanced at these redshifts, and generally towards earlier cosmic times. This shows that Mpc-scale events can have an important impact on the formation and evolution of AU-scale planetary systems. Figure from Pfeffer et al.~(2018).}
   \label{fig:emosaics}
\end{center}
\end{figure}

The E-MOSAICS simulations show that the CFE systematically increases with redshift (Pfeffer et al.~2018), such that more (proto)planetary systems are affected by external disruption at earlier cosmic epochs (see Figure~\ref{fig:emosaics} for an example). The conditions in the present-day Universe thus represent a lower limit to the fraction of (proto)planetary systems undergoing externally-driven disruption. At the formation time of the Sun, the CFE was typically a factor of 1.5--2 larger than at the present day, whereas the CFE peaked at 30--50\% at the peak of the star formation history at $z\sim2$, corresponding to a lookback time of about 10 Gyr. At lookback times $>8$~Gyr, most planetary systems are expected to be affected by environmental effects. In addition, the sharp peaks at redshifts $z=\{0.7,2,2.5\}$ correspond to galaxy mergers, which drive an increase of the gas pressure and a corresponding increase of the CFE. As a result, the external disruption of (proto)planetary systems is enhanced at these redshifts, and generally towards earlier cosmic times. This shows that Mpc-scale events can have an important impact on the formation and evolution of AU-scale planetary systems.

\section{Conclusion: planetary systems are shaped by galactic environment}
In summary, the presented results lead to the following conclusions.
\begin{enumerate}
\item
Dynamical interactions and (especially) external photoevaporation affect properties of (proto)planetary discs and planetary systems, mostly limiting the maximum radius where planets reside and shortening disc lifetimes.
\item
These environmental effects can sterilise planetary systems by preventing planet formation or ejecting planets from the habitable zone.
\item
The fraction of planetary systems potentially subject to these influences is environmentally dependent, increasing from a few \% at present (with a CFE of 7\% in the current solar neighbourhood) to $>50\%$ when most stars in Milky Way formed.
\end{enumerate}
Building on these results, we have derived a model for PPD lifetimes as a function of the local density at which star formation proceeds in the context of the galactic environment (Winter, Kruijssen, Chevance, Clarke, Keller, Longmore, to be submitted). This model will enable a systematic assessment of environmental effects on (proto)planetary systems.

\begin{discussion}

\discuss{Palous} {What is the role of metallicity in setting the cluster formation efficiency?}

\discuss{Kruijssen} {In the model for the cluster formation efficiency that I showed, metallicity could enter through the star formation efficiency or the bound fraction. However, neither of these have strong dependencies on metallicity. So metallicity is largely a tracer of birth environment (galaxy mass, redshift; see the colour coding in Figure~\ref{fig:emosaics}), but plays no role in setting the cluster formation efficiency directly.}

\end{discussion}

\end{document}